\documentclass[1p]{elsarticle}

\usepackage{hyperref}

\usepackage{amsthm,amsmath}
\RequirePackage{natbib}
\RequirePackage{hyperref}
\usepackage{amssymb}
\usepackage[utf8]{inputenc} 
\usepackage[finalizecache=false,frozencache=true,cachedir=.]{minted}
\setminted{frame=lines,autogobble}
\usepackage{algorithm}
\usepackage{algpseudocode}

\algblockdefx{MAP}{ENDMAP}[1]%
             {\textbf{map} #1}%
             {\textbf{end map}}

\algblockdefx{MAPP}{ENDMAPP}[1]%
             {\textbf{mapPartition} #1}%
             {\textbf{end mapPartition}}
\algblockdefx{REDUCE}{ENDREDUCE}[1]%
             {\textbf{reduce} #1}%
             {\textbf{end reduce}}

\begin{document}

\begin{frontmatter}

\title{DPASF: A Flink Library for Streaming Data preprocessing}


\author[mymainaddress]{Alejandro
  Alcalde-Barros\corref{correspondingauthor}}
\ead{algui91@gmail.com}
\cortext[correspondingauthor]{Corresponding author}

\author[mymainaddress]{Diego García-Gil\corref{correspondence}}
\ead{djgarcia@decsai.ugr.es}
\cortext[correspondence]{Correspondence}

\author[mymainaddress]{Salvador García}
\author[mymainaddress]{Francisco Herrera}

\address[mymainaddress]{Department of Computer Science and Artificial
  Intelligence, CITIC-UGR (Research Center on Information and
  Communications Technology), University of Granada, Calle Periodista
  Daniel Saucedo Aranda, 18071, Granada, Spain}

\begin{abstract}

Data preprocessing techniques are devoted to correct or alleviate errors in data.
Discretization and feature selection are two of the most extended data preprocessing
techniques. Although we can find many proposals for static Big Data preprocessing, there is
little research devoted to the continuous Big Data problem. Apache Flink is a recent and novel
Big Data framework, following the MapReduce paradigm, focused on distributed stream and
batch data processing.

In this paper we propose a data stream library for Big Data preprocessing, named DPASF,
under Apache Flink. We have implemented six of the most popular data preprocessing
algorithms, three for discretization and the rest for feature
selection.

The algorithms have been tested using two Big Data datasets. Experimental results
show that preprocessing can not only reduce the size of the data, but to maintain or even
improve the original accuracy in a short time.

DPASF contains useful algorithms when dealing with Big Data data streams. The
preprocessing algorithms included in the library are able to tackle Big Datasets efficiently and
to correct imperfections in the data.

\end{abstract}


\end{frontmatter}

\section{Background}

In recent years, the amount of data generated can no longer be
treated directly by humans or manual applications, there is a need to
analyze this data automatically and at large scale~\cite{Garcia2016}.
This is what is know to be Big Data, data generated at high volume,
velocity and variety. This kind of data require a new high-performance
processing and it can be found in many fields in these times~\cite{Saeed2018}.

In order to extract quality information from data, a previous steps to
learning must be performed. This is known as data
preprocessing~\cite{GARCIA20161}, this step is almost mandatory in
order to obtain a good model. Data preprocessing~\cite{Garca2015} deals
with missing values, noise, and redundant features~\cite{Garcia2016}
among others.

Data streams~\cite{RAMIREZGALLEGO201739} are sequences of unbounded and ordered data that arrives
one at a time. This imposes restrictions on the learning algorithms
which do not appear on static data. Therefore, new algorithms that can
deal with this kind of data must be developed.

To address the problem of dealing with such quantity of data in the
Big Data era, distributed frameworks such as Apache Spark~\cite{spark}
and Apache Flink~\cite{flink, friedman2016introduction} has been
developed. Apache Spark is well known, designed as a fast and general
engine for large-scale in-memory data processing. Apache Flink focus,
on the other hand, resides on distributed streams and batch data
processing~\cite{diego}. On top of that, Flink is the only system
incorporating a distributed dataflow runtime exploiting pipelined
streaming execution for both stream and batch workloads, exactly once
state consistency using a lightweight checkpoint system, iterative
processing built-in and window semantics supporting out-of-order
processing~\cite{Carbone1059537}.

This paper present the data stream library for Big Data preprocessing
named DPASF, where six data preprocessing algorithms are implemented
for Apache Flink, focusing on the discretization and feature selection
problems. The selected algorithms are InfoGain~\cite{infogain},
FCBF~\cite{fcbf} and OFS~\cite{ofs} for feature selection and
IDA~\cite{ida}, LOFD~\cite{lofd} and PiD~\cite{pid} for discretization.

The rest of the paper describes the theoretical background of the
algorithms, experimental results and a brief tutorial on how to use
them. Source code is available on
GitHub\footnote{\url{https://github.com/elbaulp/dpasf}}~\cite{dpasf}.
The algorithms chosen are the most representative in data streaming
preprocessing, in addition, have shown positive results.



\subsection{Big Data}
In general, Big Data ~\cite{Landset2015,Rao2018,RamrezGallego2018} is
known to be data that is too big or too complex to handle by
conventional tools and on a single machine. For this reason, there is
a increasing need in developing new tools that can handle all this
amount of data efficiently. In order to accomplish it, distributed
frameworks like Hadoop~\cite{hadoop}, Spark~\cite{spark} and
Flink~\cite{flink} were developed. This type of frameworks allow to
process larges amount of data in a scalable way.

One common way of defining Big Data is to describe it in terms of three
dimensions, also known as the 3 V's~\cite{ref6} (Volume, Velocity and
Variety). Volume just refers to how much data there is. Velocity refers
to the speed at which data is processed and analyzed. Lastly, Variety
refers to how many different data formats there are to be analyzed.

\subsection{Data Streaming}

The main characteristics of streaming data~\cite{joaogama} are the
following. In streaming data, instances are not available beforehand,
it becomes available in a sequence fashion, one by one, or in batches.
Instances can arrive quickly and at irregular time intervals. Due to
streaming data is unbounded, it may be infinite and can not be stored
in memory. Each instance is accessed only one time (most of the time)
and then is discarded. In order to provide real-time processing,
instances are processed within a limited amount of time. The intrinsic
characteristics of data are subject to change over time, this is what
is know as concept drift~\cite{Gama:2014:SCD:2597757.2523813}.

Concept drift~\cite{RAMIREZGALLEGO201739} is the main problem within
streaming data, as it is important for the algorithm to detect it and
update the learned model to reflect the changes underlying to the
data.

\subsection{Data preprocessing}
Before applying any data mining process, it is necessary to adapt the
data into the requirements imposed by each learning algorithm and
clean the data properly.

Although data preprocessing is a critical step, often it is time
consuming. There are two types of data preprocessing, those designed
to reduce the complexity in the data, and those designed to prepare
the data, this means data transformation, cleaning, normalizing etc.
The former is called data reduction, the latter data preparation. When
these techniques are applied, the data is in its final stage to be fed
to the data mining algorithm.

Among data transformation, feature
selection~\cite{Garcia2016,li2017feature} takes care of selecting only
relevant and non-redundant attributes. The aim of this type of data
preprocessing is to obtain a subset of the original data that stills
maintains the ability to describe the inherent concept. As a side
effect, reducing the complexity of the data also results in better
efficiency in terms of the amounts of time to learn a model, as well
as preventing over-fitting.

Discretization~\cite{ida} is a technique for reducing the complexity of the data
by dividing the domain of the variables into bins defined by cut
points. This process transform quantitative values into qualitative,
as cut points define a set of non-overlapping  intervals. Once the
algorithm has computed cut points for each attribute, data is then
mapped to its corresponding interval~\cite{Garcia2016}.

\subsection{Apache Flink}

Although it may seems Apache Spark and Apache Flink are similar, they
are designed to address different problems. Apache Spark process all
data using a batch approach, it lacks a true streaming processing.
Apache flink fills this gap. Flink provides both kind of processing,
batch and streaming, but Flink process streaming data as it happens,
in an online fashion. In other words, Spark “emulates” streaming by
processing streaming data in mini-batches, whereas Flink process them
online. This makes Flink more efficient in terms of low latency.

Apache Flink has a fault tolerance system in order to recover from
exceptions that may occur. It is designed to work at low latency even
with large amounts of data.

\section{Theoretical description of the algorithms presented}
This section presents a theoretical description of the implemented
algorithms, as well as an introduction to feature selection
(InfoGain~\cite{infogain}, FCBF~\cite{fcbf} and OFS~\cite{ofs}) and
discretization (IDA~\cite{ida}, LOFD~\cite{lofd} and PiD~\cite{pid}).

\subsection{Feature Selection}

Feature selection~\cite{ramirez2018information} is meant to reduce the
dimensionality of a dataset by removing irrelevant and redundant
features. By doing this, a subset of the original features that still
describes the inherent concept behind the data is returned. FS methods
can be divided in the following categories:

\begin{itemize}
\item \textit{Wrapper Methods} It uses an external evaluator, which
  depends on a learning algorithm.
\item \textit{Filtering Methods} It uses selection techniques based on
  separability measures or statistical dependencies.
\item \textit{Embedded Methods} It uses a search procedure implicitly
  embedded on the classifier or regressor.
\end{itemize}

In general, filtering methods tend to achieve better results when
generalizing due to learning independence. In addition, filter methods
are more efficient than wrapper methods, since the latter need to
learn a model first. Therefore, in the context of big data, filtering
methods are more widely used. Information Gain~\cite{infogain},
OFS~\cite{ofs} and FCBF~\cite{fcbf} are the most popular preprocessing
algorithms in this area.


\subsubsection{Information Gain}

This feature selection scheme, described in~\cite{infoGain} is formed by two
steps: An incremental feature ranking method, and an incremental
learning algorithm that can consider a subset of the features
during prediction.

For this algorithm, the conditional entropy with respect to the class
is computed with

\begin{equation}
  H(X|Y) = -\sum_j P(y_j)\sum_i P(x_i|y_j)\log_2(P(x_i|y_j))
\end{equation}

then, the Information Gain (IG) is computed for each attribute with

\begin{equation}IG(X|Y) = H(X) - H(X|Y)\end{equation}

Once the algorithm has all Information Gain values for each attribute,
the top \(N\) are selected as best features.

\subsubsection{Online Feature Selection (OFS)}
\label{sec:org4c854c9}
OFS~\cite{ofs} proposes an \(\epsilon\)-greedy online feature selection
method based on weights generated by an online classifier (neural
networks) which makes a trade-off between exploration and
exploitation of features.

The main idea behind this algorithm is that when a vector \(\mathbf{x}\)
falls withing a \(L1\) ball, most of its numerical values are
concentrated in its largest elements, therefore, removing the smallest
values will result in a small change in the original vector \(x\) as
measured by the \(L_q\) norm. This way, the classifier is restricted to a
\(L1\) ball:

\begin{equation}
  \Delta_R = \left\{ \mathbf{w} \in \textsc{R}^d : ||\mathbf{w}||_1 \leq
  \textsc{R} \right\} \end{equation}

OFS maintains an online classifier \(\mathbf{w}_t\) with at most \(B\) nonzero
elements. When an instance \((\mathbf{x}_t, y_t)\) is incorrectly
classified, the classifier gets updated through online gradient
descent and then it is projected to a \(L2\) ball to delimit the
classifier norm. If the resulting classifier \(\hat{\mathbf{w}}_{t+1}\) has more than
\(B\) nonzero elements, the elements with the largest absolute value
will be kept in \(\hat{\mathbf{w}}_{t+1}\).

The above approach presents an inefficiency, even although the
classifier consists in \(B\) nonzero elements, full knowledge of the
instances is required, that is, each attribute \(\mathbf{x}_t\) must be
measured and computed. As a solution, OFS limit online feature selection to no
more than \(B\) attributes of \(\mathbf{x}_t\)

\subsubsection{Fast Correlation-Based Filter (FCBF)}
\label{sec:orgdeb22c0}
FCBF~\cite{fcbf} is a multivariate feature selection method where the class
relevance and the dependency between each feature pair are taken into
account. Based on information theory, FCBF uses symmetrical
uncertainty to calculate dependencies of features and the class
relevance. Starting with the full feature set, FCBF heuristically
applies a backward selection technique with a sequential search
strategy to remove irrelevant and redundant features. The algorithm
stops when there are no features left to eliminate.

The algorithm chooses as a correlation measure the entropy of a
variable \emph{\(X\)}, which is defined as
\begin{equation}H(X) = -\sum_i P(x_i)\log P(x_i)\end{equation}
and the entropy of \emph{\(X\)} after observing values of another variable
\(Y\) is defined as
\begin{equation}H(X|Y) = -\sum_j P(y_j)\sum_i P(x_i|y_j)\log_2(P(x_i|y_j))\end{equation}
where \(P(x_i)\) is the prior probability for all values of \(X\) and
\(P(x_i|y_j)\) is the posterior probability of \(X\) given the values of
\(Y\). With this information, a measure called \emph{Information Gain} can be
defined:
\begin{equation}IG(X|Y) = H(X) - H(X|Y)\end{equation}
According to IG, a feature \(Y\) is more correlated to \(X\) than to a
feature \(Z\) if \(IG(X|Y) > IG(Z|Y)\).

Now is all set to define the main measure for FCBF, \emph{symmetrical
  uncertainty}~\cite{su}. As a pre-requisite, data must be normalized in order to
be comparable.
\begin{equation}
  SU(X, Y) = 2\left [\frac{IG(X|Y)}{H(X) + H(Y)} \right ]
\end{equation}
\emph{SU} compensate the bias in \emph{IG} toward features with more values and
normalizes its values to the range \([0,1]\). A \emph{SU} value of 1
indicates total correlation whereas a value of 0 indicates
independence.

The algorithm follows a two step approach, first, it has to decide if
a feature is \emph{relevant} to the class and two, decide if those features
are \emph{redundant} with respect to each other.

To solve the first step, a user defined \emph{SU} threshold can be defined.
If \(SU_{i,c}\) is the \emph{SU} value for feature \(F_i\) with the class \(c\),
the subset \(S'\) of relevant features can be defined with a threshold
\(\delta\) such that \(\forall F_i \in S', 1 \leq i \leq N, SU_{i,c} \geq
\delta\).

For the second step, in order to avoid analysis of pairwise
correlations between all features, a method to decide whether the
level of correlation between two features in \(S'\) is high enough to
produce redundancy is needed in order to remove one of them. Examining
the value \(SU_{j,i} \forall F_j \in S' (j \neq i)\) allow to estimate
the level to what \(F_j\) is correlated by the rest of features in \(S'\).

The last piece of the algorithm comprehend two definitions:

\emph{Definition 1} (Predominant Correlation). The correlation between a
feature \(F_i\) and the class \(C\) is predominant \emph{iff} \(SU_{i,c} \geq
\delta\) and \(\forall F_j \in S' (j \neq i) \nexists F_j\) such that
\(SU_{j,i} \geq SU_{i,c}\)

If such feature \(F_j\) exists to a feature \(F_i\), it is called a
redundant peer to \(F_i\) and its added to a set \(S_{P_i}\) identifying
all the redundant peers for \(F_i\). \(S_{P_i}\) is divided in two parts:
\(S_{P_i}^+\) and \(S_{P_i}^-\), where \(S_{P_i}^+\) \(= \{F_j|F_j\in S_{P_i},
SU_{j,c} > SU_{i,c}\}\) and \(S_{P_i}^-\) \(= \{F_j|F_j\in S_{P_i},
SU_{j,c} \leq SU_{i,c}\}\)

\emph{Definition 2} (Predominant Feature). A feature is predominant to the
class \emph{iff} its correlation to the class is predominant or can become
predominant after removing all its redundant peers.

With the definitions above, a feature will be a \emph{good feature} if it
is \emph{predominant} in predicting the class. This two definitions along
with the next heuristics can effectively identify predominant features
and remove the need of pairwise comparisons.

\emph{Heuristic 1} (when \(S_{P_i}^+\) \(= \emptyset\)). \(F_i\) is a
predominant feature, delete all features in \(S_{P_i}^-\) and stop
searching for redundant peers for those features.

\emph{Heuristic 2} (when \(S_{P_i}^+\) \(\neq \emptyset\)). All features in
\(S_{P_i}^+\) are processed before making decisions on \(F_i\). If none
of them become predominant go to Heuristic 1, else remove \(F_i\) and
decide if features in \(S_{P_i}^-\) need to be removed based on other
features in \(S'\).

\emph{Heuristic 3} (Start point). The algorithm begins examining the
feature with the largest \(SU_{i,c}\), as this feature is always
predominant and acts as a starting point to remove redundant features.

\subsection{Discretization}

Broadly speaking, discretization~\cite{ramirez2016data} translates
quantitative data into qualitative data, trying to avoid an overlap
between the continuous domain of the variable. This process results in
a mapping of a value to a given interval. For this reason,
discretization can be considered as a data reduction process, since it
reduces data from a numerical domain to a subset of categorical
values.

More formally, a discretization \(\delta\) of a numeric attribute \emph{\(X_i\)} is a set of
\emph{m} intervals called \emph{bins}. The bins are defined by cut points \(\{
k_1, \dots, k_{m-1}\}\) that divide the domain of \emph{\(X_i\)} into \emph{m}
bins where \(b_1 = \left [-\infty, k_1\right ], b_m = \left [k_{m-1},
  \infty\right ]\) and for \(1 < i < m, b_i = \left ( k_{i-1}, k_i \right
]\). Therefore, a discretization for an attribute \emph{\(X_i\)} is a mapping
  between values \emph{\(v\)} of \emph{\(X_i\)} and its bin indexes \(\delta v = z\)
  such that \(v\in b_z\).

The most popular discretization algorithms are IDA~\cite{ida},
PiD~\cite{pid} and LOFD~\cite{lofd}.

\subsubsection{Incremental Discretization Algorithm (IDA)}

IDA~\cite{ida} approximates quantile-based discretization on the entire
data stream encountered to date by maintaining a random sample of the
data which is used to calculate the cut points. IDA uses the reservoir
sampling algorithm to maintain a sample drawn uniformly at random from
the entire stream up until the current time.

  In IDA, a random sample is used because it's not feasible for
  high-throughput streams to maintain a complete record of all the
  values seen so far. The sample method used is called reservoir
  sampling~\cite{reservoir}, and mantains a random sample of \emph{s} values
  \(V_i\) for each attribute \emph{\(X_i\)}. The first \emph{s} values that
  arrives for each \emph{\(X_i\)} are added to its corresponding \(V_i\).
  Thereafter, every time a new instance \(\left \langle \mathbf{x}_n, y_n \right
  \rangle\) arrives, each of its values \(\mathbf{x}_n^i\) replaces a randomly
  selected value of the corresponding \(V_i\) with probability \(s/n\).

  Each value of each attribute is stored in a vector of interval heaps~\cite{ih}. \(V_i^j\) stores the values for the j\(^{\text{th}}\) bin of \emph{\(X_i\)}.
  The reason to use a Interval Heap is that it provides efficient access
  to minimum and maximum values in the heap and direct access to random
  elements within the heap.

  \subsubsection{Partition Incremental Discretization algorithm (PiD)}
  \label{sec:org29be65f}
  PiD~\cite{pid} performs incremental discretization. The basic idea
  is to perform the task in two layers. The first layer receives the
  sequence of input data and keeps some statistics on the data using
  many more intervals than required. Based on the statistics stored by
  the first layer, the second layer creates the final discretization.
  The proposed architecture processes streaming exam ples in a single
  scan, in constant time and space even for infinite sequences of
  examples.

  PiD~\cite{pid} performs incremental discretization. The basic idea
  is to perform the task in two layers.

  The first layer receives the sequence of input data and the range of
  the variable and keeps some statistics on the data using many more
  intervals than required. The range of the variable is used to
  initialize the cut points with the same width. Each time a new value
  arrives, this layer is updated in order to compute the corresponding
  interval for the value. Each interval has a internal count of the
  values it has seen so far. When a counter for an interval reach a
  threshold a split process is triggered to generate two new intervals.
  If the interval triggering the split process is the last or the first,
  a new interval with the same step is created. Otherwise the interval
  is splitted in two. In summary, the first layer simplifies and
  summarizes the data.

  Based on the statistics stored by the first layer, the second layer
  creates the final discretization. The proposed architecture processes
  streaming examples in a single scan, in constant time and space even
  for infinite sequences of examples. To accomplish it, this layer
  merges the set of intervals computed in the previous layer.

  PiD stores the information about the number of examples per class in
  each interval in a matrix. In this matrix, columns corresponds with
  the number of intervals and rows with the number of classes. With this
  information, the conditional probability of an attribute belonging to
  an interval given that the corresponding example belongs to a class
  can be computed as \(P(b_i < x \leq b_{i+1} | Class_j)\).

  To perform the actual discretization \emph{Recursive entropy
    discretization}~\cite{fayyad1992} is used. This algorithm was
  developed by Fayyad and Irani~\cite{fayyadmld}. It uses the class
  information entropy of two candidate partitions to select the
  boundaries for discretization. It begins searching for a single
  threshold that minimizes the entropy over all possible cut points,
  then, it is applied recursively to both partitions. It uses
  the \emph{minimum description length}~\cite{mdl} principle as stop
  criteria. The algorithm works as follow:

  First, the entropy before and after the split is computed as well as
  its information gain. Then, the entropy for both left and right splits
  is computed and finally the algorithm check if the split is accepted
  with the following formula

  \begin{equation}
    Gain(A,T;S) < \frac{\log_2 (N-1)}{N} + \frac{\Delta (A, T;S)}{N}
  \end{equation}

  where \(N\) is the number of instances in the set \(S\),

  \begin{equation}
    Gain(A,T;S) = H(S) - H(A,T;S)
  \end{equation}
  and
  \begin{equation}
    \Delta(A,T;S) = \log_2 (3^k - 2) - \left [ k\cdot H(S) - k_1 \cdot H(S_1) - k_2
      \cdot H(S_2) \right ]
  \end{equation}
  where k\(_{\text{i}}\) is the number of class labels represented in the set S\(_{\text{i}}\).

  \subsubsection{Local Online Fusion Discretizer (LOFD)}
  \label{sec:orgf97de91}
  LOFD~\cite{lofd} is an online, self-adaptive discretizer for
  streaming classification. It smoothly adapts its interval limits
  reducing the negative impact of shifts and analyze interval
  labeling and interaction problems in data streaming. Interaction
  discretizer-learner is addressed by providing 2 alike solutions.
  The algorithm generates an online and self-adaptive discretization
  solution for streaming classification which aims at reducing the
  negative impact of fluctuations in evolving intervals.

  The algorithm is constituted by two phases, the main process, at
  instance level, and the merge/split process, at interval level. The
  main process works as follows. First, discrete interval are
  initialized following the static process defined in~\cite{fusinter}.
  The discretization is then performed on the first \emph{initTh} instances.
  From that moment on, LOFD updates the scheme of intervals in each
  iteration and for each attribute. For each new instance, it retrieves
  its ceiling interval (implemented as a red-black tree). If the point
  is above the upper limit a new interval is generated at that point,
  being that point the new maximum for the current attribute. A merge
  between the old last interval and the new is evaluated by computing
  the quadratic entropy, if the result is lower than the sum of both
  parts, the merge is accepted.

  Finally, each point is added to a queue with timestamp to control
  future removals in the case the histogram overflows. If necessary,
  LOFD recovers points from the queue in ascending order and remove
  them until there is space left in the histogram.

  The split/merge phase is triggered each time a boundary point is
  processed. The new boundary point splits an interval in two, one
  interval contains the points in the histogram with values less than or
  equal to the new point and keeps the same label. Each time a new
  interval is generated, the merge process is triggered for the
  intervals being divided and its neighbors.

  \section{Implementation}

  This section presents the pseudocode for the implemented algorithms. The
  following Flink primitives have been used:

  \begin{itemize}
  \item \textbf{map} The Map transformation applies a user-defined map function on each element of a DataSet
  \item \textbf{reduce} A Reduce transformation reduces the dataset to a
    single element using a user-defined reduce function.
  \item \textbf{mapPartition} MapPartition transforms a parallel partition in a single function call.
  \item \textbf{reduceGroup} A GroupReduce transformation that is applied on a grouped DataSet calls a user-defined group-reduce function for each group. The difference between this and Reduce is that the user defined function gets the whole group at once.
  \end{itemize}

  Algorithm~\ref{alg:fcbf} shows pseudocode for FCBF, the SU value is
  computed for each attribute in parallel. All SU values are then
  filtered according to the threshold parameter and then sorted
  descendingly. With this final sorted values, FCBF algorithm is applied
  like originally described in~\cite{fcbf}. Algorithm~\ref{alg:su} shows
  how Symmetrical Uncertainty is computed in a distributed fashion.
  First, each parallel partition compute the partial counts of each
  value, then this partial counts are aggregated using a reduce function
  in order to compute the total counts. With this information,
  probabilities for each value are computed and its entropy and mutual information are
  calculated. Finally, it returns the corresponding SU value for that attribute.

  \begin{algorithm}[!t]
    \floatname{algorithm}{Algorithm}
    \caption{FCBF Algorithm}
    \label{alg:fcbf}
    \begin{algorithmic}[1]
      \State \textbf{Input:} \textit{data} a DataSet LabeledVector (label, features)
      \State \textbf{Input:} \textit{thr} threshold
      \State \textbf{Output:} DataSet with the most important features
      \State $su \gets$ \For{$i \gets 0$ until $nAttrs$}
      \State $attr \gets$ \MAP $instance \in data$
      \State $(label, feature_i)$ \ENDMAP
      \State \textbf{yield} \Call{SU}{$attr$}
      \EndFor
      \State $suSorted \gets$ \Call{filter}{su >
        thr}.\Call{SortDesc}{}
      \State $sBest \gets$ \Call{FCBF}{suSorted}
      \State \Return $sBest$
    \end{algorithmic}
  \end{algorithm}

  \begin{algorithm}[!t]
    \floatname{algorithm}{Algorithm}
    \caption{Symmetrical Uncertainty function (SU)}
    \label{alg:su}
    \begin{algorithmic}[1]
      \State \textbf{Input:} \textit{attr} Attribute to comute SU to
      \State \textbf{Output:} SU value for \textit{attr}
      \State $xypartialCounts \gets$ \MAPP
      $(y, x) \in attr$
      \State $xPartialCounts \gets$ \Call{computeCounts}{x}
      \State $yPartialCounts \gets$ \Call{computeCounts}{y}
      \State $(xPartialCounts, yPartialCounts)$
      \ENDMAPP
      \State $totalCounts \gets$ \Call{reduce}{xypartialCounts}
      \State $su \gets$
      \MAP $(xcounts, ycounts, x, y) \in totalCounts$
      \State $px \gets$ \Call{prob}{x}
      \State $py \gets$ \Call{prob}{y}
      \State $hx \gets$ \Call{entropy}{x}
      \State $hy \gets$ \Call{entropy}{y}
      \State $mu \gets$ \Call{mutualInformation}{x,y}
      \State $\frac{2mu}{hx + hy}$
      \ENDMAP
    \end{algorithmic}
  \end{algorithm}

  Algorithm~\ref{alg:ig} shows the implementation for Information Gain~\cite{infoGain}.
  First the frequencies of each value with respect to the class label
  are computed. With this information, the total entropy of the dataset
  is computed. Next, for each attribute, its frequency, probability,
  entropy and conditional entropy are computed. Finally, the information
  gain for the i-th attribute its computed and stored into
  \textit{gains}. Algorithm~\ref{alg:freq} shows how frequencies are
  computed.

  \begin{algorithm}[!t]
    \floatname{algorithm}{Algorithm}
    \caption{InfoGain Algorithm}
    \label{alg:ig}
    \begin{algorithmic}[1]
      \State \textbf{Input:} \textit{data} a DataSet LabeledVector (label, features)
      \State \textbf{Input:} \textit{selectNF} Number of features to select
      \State \textbf{Output:} DataSet with the most \textit{selectNF} important features
      \State $freqs \gets$ \Call{frequencies}{data, groupBy label}
      \State $H \gets$ \Call{Entropy}{freqs}
      \State $gains \gets$ \MAP $i \in 0$ until $nFeatures$
      \State $freqs \gets$ \Call{frequencies}{data, $feature_i$}
      \State $px \gets$ \Call{probs}{freqs}
      \State $H \gets$ \Call{entropy}{freqs}
      \State $H(Y|Feature_i) \gets$ \Call{ConditionalEntropy}{freqs}
      \State $H - H(Y|Feature_i)$
      \ENDMAP
      \State \Return \Call{selectFeatures}{selectNF, gains}
    \end{algorithmic}
  \end{algorithm}

  \begin{algorithm}[!t]
    \floatname{algorithm}{Algorithm}
    \caption{Frequencies function}
    \label{alg:freq}
    \begin{algorithmic}[1]
      \State \textbf{Input:} \textit{attr} attribute to compute
      frequencies to
      \State \textbf{Input:} \textit{f} function to group by
      \State \textbf{Output:} Frequencies for \textit{attr} using \textit{f}
      \State $grouped \gets groupBy(data, f)$
      \State $freqs \gets reduceGroup(grouped)$
      \State \Return $freqs$
    \end{algorithmic}
  \end{algorithm}

  Algorithm~\ref{alg:ofs} shows the pseudocode for OFS~\cite{ofs}, this
  algorithm maps each label and feature with its corresponding value for
  the original OFS algorithm.

  \begin{algorithm}[!t]
    \floatname{algorithm}{Algorithm}
    \caption{OFS Algorithm}
    \label{alg:ofs}
    \begin{algorithmic}[1]
      \State \textbf{Input:} \textit{data} a DataSet LabeledVector (label, features)
      \State \textbf{Input:} \textit{$\eta$}  parameter
      \State \textbf{Input:} \textit{$\lambda$} parameter
      \State \textbf{Input:} \textit{selectNF} Number of features to select
      \State \textbf{Output:} DataSet with the most \textit{selectNF} important features

      \State $finalweights \gets$
      \MAP $(label, features) \in data$
      \State \Call{OFS}{label, features}
      \ENDMAP

    \end{algorithmic}
  \end{algorithm}

  Algorithm~\ref{alg:ida} shows pseudocode for IDA~\cite{ida}. This
  algorithm first compute the cut points for the dataset with the
  desired number of bins. In order to compute the cut points, each
  instance is mapped to the result of IDA, which returns the computed
  cut points. To achieve it, each feature is zipped with its index,
  and then folded with its corresponding class label and a zero
  feature vector that will be filled in each iteration of the fold
  operation, with the returned value of IDA algorithm. Once cut points
  are stored, line 5 in Algorithm~\ref{alg:ida} discretizes the data
  according to those cut points.

  \begin{algorithm}[!t]
    \floatname{algorithm}{Algorithm}
    \caption{IDA Algorithm}
    \label{alg:ida}
    \begin{algorithmic}[1]
      \State \textbf{Input:} \textit{data} a DataSet LabeledVector (label, features)
      \State \textbf{Input:} \textit{bins} number of bins
      \State \textbf{Output:} Discretized dataset with desired number
      of \textit{bins}
      \State $cuts\gets$ \MAP $((y, x) \in data)$
      \State $zipped\gets$ \Call{zipWithIndex}{x}
      \State \Call{$FoldLeft((y, empty feature))$}{$IDA()$}
      \ENDMAP
      \State \Return \Call{discretize}{data, cuts}
    \end{algorithmic}
  \end{algorithm}

  Algorithm~\ref{alg:pid} shows pseudocode for PiD~\cite{pid}, this
  algorithm first initialize the required data structures using a map
  function, this map function expands the dataset and adds to it a
  histogram and a counter of total instances seen so far. Then this data
  is reduced computing in each reduce step the layers one and two as
  described in the original algorithm~\cite{pid}. Once this reduce stage
  has been completed, it returns the discretized data using the
  previously computed cut points.

  \begin{algorithm}[!t]
    \floatname{algorithm}{Algorithm}
    \caption{PiD Algorithm}
    \label{alg:pid}
    \begin{algorithmic}[1]
      \State \textbf{Input:} \textit{data} a DataSet LabeledVector (label, features)
      \State \textbf{Input:} \textit{$\alpha$} parameter
      \State \textbf{Input:} \textit{step} parameter
      \State \textbf{Output:} Discretized dataset
      \State $cuts\gets$ \MAP $instance \in data$
      \State $(instance, Histogram, 1)$
      \ENDMAP
      \REDUCE $(m1, m2) \in cuts$
      \State \Call{UpdateLayer1}{m1, m2}
      \State \Call{UpdateLayer2}{m1, m2}
      \ENDREDUCE
      \State \Return \Call{discretize}{data, cuts}
    \end{algorithmic}
  \end{algorithm}

  Algorithm~\ref{alg:lofd} shows pseudocode for LOFD~\cite{lofd}. This
  algorithm first instantiate a LOFD helper, and maps the data to the
  computed cut points this helper returns. Once all cutpoints has been
  collected, the reduce function extract only the most recently computed
  cut points and apply the discretization based on them.

  \begin{algorithm}[!t]
    \floatname{algorithm}{Algorithm}
    \caption{LOFD Algorithm}
    \label{alg:lofd}
    \begin{algorithmic}[1]
      \State \textbf{Input:} \textit{data} a DataSet LabeledVector (label, features)
      \State \textbf{Output:} Discretized dataset
      \State $lofd\gets LOFDInstance$
      \State $cuts\gets$ \MAP $x \in data$
      \State $discretized \gets lofd.applyDiscretization(x)$
      \For{s in 0 until discretized.size}
      \State $lofd.getCutpoints(s)$
      \EndFor
      \ENDMAP
      \REDUCE $(\_, b) \in cuts$
      \State $b$
      \ENDREDUCE
      \State \Return \Call{discretize}{data, cuts}
    \end{algorithmic}
  \end{algorithm}

  \section{Results}

  This section present in detail the six algorithms implemented in
  Apache Flink, there are three algorithms for feature selection and three for
  discretization.

  \subsection{Examples}
  The software has been implemented in the Scala programming
  language\footnote{\url{https://scala-lang.org/}} language. As
  mentioned above, DPASF consist of six algorithms for data
  streams, three discretization methods and three feature
  selection methods. The software can be found on
  GitHub\footnote{\url{https://github.com/elbaulp/DPASF}}. The next
  section presents how to use each algorithm within Apache Flink.

  \subsection{Usage}
  \subsubsection{Feature Selection}
  \paragraph{FCBF}

  In order to benefit from the Apache Flink framework, symmetrical uncertainty computation
  for each pair of attributes are distributed across each node in order
  to speed the process.

  Suppose the data set to be used is the Abalone DataSet\footnote{\url{https://archive.ics.uci.edu/ml/datasets/Abalone}}, to
  load it in Apache Flink:
  \begin{minted}[mathescape=true]{scala}
    val abaloneDat = env.readCsvFile[(Int, Double,..., Int)]
    (getClass.getResource("/abalone.csv").getPath)
    .name("Reading Abalone DS")
    val abaloneDS = abaloneDat
    .map { tuple =>
      val list = tuple.productIterator.toList
      val numList = list map { x =>
        x match {
          case d: Double => d
          case i: Int => i
        }
      }
      LabeledVector(numList(8), DenseVector(numList.take(8).toArray))
    }.name("Abalone DS")
  \end{minted}
  Then, a \texttt{FCBFTransformer} must be instantiated, configure its
  parameters and finally define a pipeline:
  \begin{minted}[mathescape=true]{scala}
    val fcbf = FCBFTransformer()
    .setThreshold(.05)

    fcbf fit abaloneDS
    val bestFeatures = fcbf transform abaloneDS
  \end{minted}

  After fitting the algorithm, calling \texttt{transform} on \texttt{fcbf} will
  return the Abalone data set with the most important features.

  \paragraph{InfoGain}

  For this algorithm, each attribute's Information Gain value is
  computed in parallel.

  The use of \texttt{InfoGainTransformer} is similar:
  \begin{minted}[mathescape=true]{scala}
    val data = Vector(
    Vector("1", "0", "10"),
    Vector("0", "0", "10"),
    Vector("1", "0", "10"),
    Vector("0", "1", "20"),
    Vector("0", "0", "10"),
    Vector("1", "1", "20"),
    Vector("1", "0", "10"))

    val gain = InfoGainTransformer()
    .setNFeatures(2)
    .setSelectNF(1)

    gain fit dataSet

    val result = gain transform dataSet
  \end{minted}
  \paragraph{OFS}

  One difference of OFS with respect to the previous algorithms is
  that it does not require a fitting phase:

  \begin{minted}[mathescape=true]{scala}
    val ofs = OFSGDTransformer()
    .setNFeature(5)
    val result = ofs transform data
  \end{minted}
  \subsubsection{Discretization}

  In this section the usage of discretization method is presented, all
  of them use the
  Iris\footnote{\url{https://archive.ics.uci.edu/ml/datasets/Iris/}}
  DataSet, loaded as:

  \begin{minted}[mathescape=true]{scala}
    // Iris POJO
    case class Iris(
    SepalLength: Double,
    SepalWidth: Double,
    PetalLength: Double,
    PetalWidth: Double,
    Class: Double)

    val data = env.readCsvFile[Iris](getClass.getResource("/iris.dat").getPath)
    val dataSet = data map { tuple =>
      val list = tuple.productIterator.toList
      val numList = list map (_.asInstanceOf[Double])
      LabeledVector(numList(4), DenseVector(numList.take(4).toArray))
    }
  \end{minted}
  \paragraph{IDA}
  For IDA, cut points are computed in parallel, in order to get the
  most recent computed cut point, data is reduced to get the latest
  set of cuts.
  \begin{minted}[mathescape=true]{scala}
    val ida = IDADiscretizerTransformer()
    .setBins(5)
    val discretizedIris = ida transform dataSet
  \end{minted}
  \paragraph{PiD}

  For PiD, the needed histogram is shared across all nodes. After
  histogram is initialized, data is reduced in order to produce the
  final histogram, where the cut points to perform the discretization are.

  In PiD, data must be normalized as a previous step, so a
  \texttt{ChainTransformer} is used in the pipeline.
  \begin{minted}[mathescape=true]{scala}
    val pid = PIDiscretizerTransformer()
    .setAlpha(.10)
    .setUpdateExamples(50)
    .setL1Bins(5)

    val scaler = MinMaxScaler()
    val pipeline = scaler.chainTransformer(pid)

    pipeline fit dataSet
    val result = pipeline transform dataSet
  \end{minted}
  \paragraph{LOFD}

  For LOFD, PiD-like approach is used. First, all features are mapped in
  order to extract the necessary information from them, then, data is
  reduced to extract the final cut points to perform discretization.

  \begin{minted}[mathescape=true]{scala}
    val lofd = LOFDiscretizerTransformer()
    .setInitTh(1)
    val discretized = ofs transform (dataSet)
  \end{minted}

  \subsection{Results}
  The experimental set up has used two datasets, \textit{ht\_Sensor}
  and \textit{skin\_nonskin}, Table~\ref{table:ds} describe them.

  \begin{table}[h!]
    \caption{Information about DataSets for experiments}
    \label{table:ds}
    \begin{tabular}{l|r|r|r}
      DataSet & Instances & Attributes & Classes\\
      \hline
      ht\_sensor & 929000 & 11 & 3\\
      skin\_nonskin & 245000 & 3 & 2\\
    \end{tabular}
  \end{table}

  All algorithms has been tested with KNN and Decision Trees using
  5-fold cross validation. A baseline is fitted without any
  preprocessing step, and other with the corresponding preprocessing
  algorithm. In addiction KNN has been fitted with k = 3 and k = 5.

  For feature selection methods, all of them have been set up to select
  50\% of features.

  Table~\ref{table:times} shows the amount of time it took to
  preprocess the data. The worst algorithm by far is IDA, which took
  about 5 hours to finish for ht\_sensor. On the contrary, the fastest
  was InfoGain. OFS could not be measured as it only accepts binary
  datasets. It is worth mentioning that these experiments could not
  have been possible in normal environments due to the amount of time
  they would have taken.

  For all experiments we have used a cluster composed of 14 computing
  nodes. The nodes hold the following characteristics: 2 x Intel Core
  i7-4930K, 6 cores each, 3.40 GHz, 12 MB cache, 4 TB HDD, 64 GB RAM.
  Regarding software, we have used the following configuration: Apache
  Flink 1.6.0, 238 TaskManagers (17 TaskManagers/core), 49 GB
  RAM/node.

  \begin{table}[h!]
    \caption{Times in seconds}
    \label{table:times}
    \centering
    \begin{tabular}{l|r|r}
      Preprocessing algorithm & ht\_sensor & skin\_nonskin\\
      \hline
      FCBF & 19 & 1\\
      OFS & - & 1\\
      InfoGain & 16 & 2\\
      IDA & 20854 & 93\\
      LOFD & 28 & 3\\
      PiD & 118 & 7\\
    \end{tabular}
  \end{table}

  Tables~\ref{table:k3} and \ref{table:k5} show the accuracy obtained
  by the algoritms, as well as the accuracy without any preprocessing.
  The three feature selection methods obtain considerable results,
  even when they are configured to remove half the features on the
  datasets. FCBF is not among the bests, but it is among the fastest,
  this may be due to the fact that it avoids computing pairwise
  comparisons when selecting features. Also, FCBF can not be set to
  select a fixed number of features, it uses a threshold to select
  features based on its SU value, so in some cases it will select less
  than half the features. InfoGain also gives excellent results, close
  to baseline and even improve when k=3. Among the discretizers, PiD
  outperforms baseline, in all cases but skin-nonsking with k=3.

  \begin{table}[h!]
    \caption{Accuracy for KNN with k = 3}
    \label{table:k3}
    \begin{tabular}{l|r|r}
      k = 3 & ht\_sensor & skin\_nonskin\\
      \hline
      No-PP & 0.9998 & \textbf{0.9995}\\
      FCBF & 0.8965 & 0.8642\\
      OFS & - & 0.8985\\
      InfoGain & \textbf{0.9999} & 0.9825\\
      IDA & 0.8845 & 0.6591\\
      LOFD & 0.9662 & 0.9755\\
      PiD & \textbf{0.9999} & 0.9966\\
    \end{tabular}
  \end{table}

  \begin{table}[h!]
    \caption{Accuracy for KNN with k = 5}
    \label{table:k5}
    \begin{tabular}{l|r|r}
      k = 5 & ht\_sensor & skin\_nonskin\\
      \hline
      No-PP & 0.9999 & 0.9994\\
      FCBF & 0.8037 & 0.8684\\
      OFS & - & 0.9006\\
      InfoGain & 0.9991 & 0.9838\\
      IDA & 0.8850 & 0.7092\\
      LOFD & 0.9665 & 0.9766\\
      PiD & \textbf{0.9999} & \textbf{0.9994}\\
    \end{tabular}
  \end{table}

  Table~\ref{table:dt} shows accuracy for a Decision Tree model,
  results are consistent with the previous model. In general, feature
  selection methods result in a decrease on accuracy whereas
  discretization methods are consistent with baseline, albeit PiD
  improves accuracy.

  \begin{table}[h!]
    \caption{Accuracy for Decision Trees}
    \label{table:dt}
    \begin{tabular}{l|r|r}
      DT & ht\_sensor & skin\_nonskin\\
      \hline
      No-PP & 70.13 & 98.45\\
      FCBF & 57.50 & 88.00\\
      OFS & - & 88.15\\
      InfoGain & 67.51 & 97.10\\
      IDA & 68.35 & 94.24\\
      LOFD & 69.85 & 94.18\\
      PiD & \textbf{71.06} & \textbf{98.74}\\
    \end{tabular}
  \end{table}
  \section{Conclusions}

  In this paper we have tackled the Big Data streaming preprocessing
  problem. We have proposed a library for Big Data data stream
  preprocessing, named DPASF, implemented in the Big Data streaming
  framework Apache Flink. This library includes six classic data
  preprocessing algorithms, three for performing discretization, and
  another three for the feature selection task. All the algorithms
  have been redesigned in order to make them able to cope with big
  datasets.

The performance of the six algorithms in Big Data scenarios has been
analyzed using two Big Data datasets. Experimental results have shown
that preprocessing can improve the original accuracy in a short amount
of time. We have also observed that choosing the right technique is
crucial depending on the problem and the classifier used.

\bibliography{mybibfile}

\end{document}